\documentstyle[aps,prc,epsfig]{revtex}
\newcommand{\beq}{\begin{equation}}
\newcommand{\eeq}{\end{equation}}
\newcommand{\bea}{\begin{eqnarray}}
\newcommand{\eea}{\end{eqnarray}}
\def\bfg{\begin{figure}}
\def\efg{  \end{figure}}
\newcommand{\bce}{\begin{center}}
\newcommand{\ece}{\end{center}}
\newcommand{\eg}{{\it e.g.}}
\newcommand{\ie}{{\it i.e.}}
\newcommand{\etal}{{\it et al.}}
\def\lsim{\mathrel{\rlap{\lower4pt\hbox{\hskip1pt$\sim$}}
    \raise1pt\hbox{$<$}}}         %less than or approx. symbol
\def\gsim{\mathrel{\rlap{\lower4pt\hbox{\hskip1pt$\sim$}}
    \raise1pt\hbox{$>$}}}         %greater than or approx. symbol
\begin{document}
\twocolumn[\hsize\textwidth\columnwidth\hsize\csname @twocolumnfalse\endcsname
%\preprint{SUNY-NTG-00-xx}
%
%
\title{Hadro-Chemistry and Evolution of (Anti-) Baryon Densities at RHIC}
 
\author{Ralf Rapp}
 
\address
{Department of Physics and Astronomy, State University of New York, 
    Stony Brook, NY 11794-3800, U.S.A.}

\date{\today} 

\maketitle
 
\begin{abstract}
The consequences of hadro-chemical freezeout for the subsequent
hadron gas evolution in central heavy-ion collisions at RHIC and LHC 
energies are discussed with special emphasis on effects due to 
antibaryons. Contrary to naive expectations, their individual 
conservation, as implied by experimental data, has significant impact on 
the chemical off-equilibrium composition of hadronic matter
at collider energies. This may reflect on a variety of observables 
including source sizes and dilepton spectra.  
\end{abstract} 
\vspace{0.58cm}
]
\begin{narrowtext}
\newpage

%%%%%%%%%%%%%%%%%%%%%%%%%%%%%%%%%%%%%%%%%%%%%%%%%%%%%%%%%%%%%%%%%%%%%%%%
%\section{Introduction}
%%%%%%%%%%%%%%%%%%%%%%%%%%%%%%%%%%%%%%%%%%%%%%%%%%%%%%%%%%%%%%%%%%%%%%%%

%The dynamics of baryon stopping and/or formation is one of the 
%intriguing questions in high-energy nuclear collisions.  
%Whereas proton-nucleus reactions can be used for a systematic 
%study of the effects arising from primordial $N$-$N$ collisions, 
%additional effects can be expected in sufficiently central collisions
%of heavy nuclei. E.g., in the case of Quark-Gluon Plasma formation, 
%hadronization is significantly delayed, and both 
%thermal and chemical equilibration,  may, in principle, 
%erase the memory on (some of the) initial state effects.  

The systematic study of hadron production in central collisions
of heavy nuclei at (ultra-) relativistic energies has revealed  
strong evidence for a high degree of equilibration 
achieved in the course of these reactions. In particular, it has 
been shown that, over a large range of collision energies 
(from SIS to SPS)~\cite{pbm95,pbm96-99,YG99,RL99,CR00,Beca01},  
the measured final state hadron abundances can be 
well described by essentially two parameters, the temperature $T$
and baryon chemical potential $\mu_B$ (which bear  
little sensitivity on centrality~\cite{CKW02}).   
Importantly, the inferred collision-energy dependence 
reinforces confidence in the notion that one is actually probing 
different regions of the QCD phase diagram in these experiments. 
The analyses are in line with the naive expectation that with 
increasing CM-energy $\sqrt{s}$ of 
the colliding nuclei, the central rapidity regions are  
increasingly governed by interactions between low-$x$ partons
of the incoming nucleons, thus
leading to a decrease in the net baryon content. 
This is reflected in  
a systematic decrease of the baryon chemical potential with 
$\sqrt s$, accompanied by an increase in temperature. 
In fact, for the top SPS energies ($\sqrt s=17$~AGeV) and for the
first RHIC data at $\sqrt{s}=130$~AGeV, the such extracted $(\mu,T)$ 
values are very close to the expected phase boundary to the 
Quark-Gluon Plasma (QGP). 
Under RHIC conditions this
implies a copious production of baryon-antibaryon pairs. Whether those
are solely of thermal origin, or what exactly the underlying microscopic
(partonic) production mechanism is, remains a matter of debate at present. 
It is clear, however, that, as with any hadronic observable, 
the impact of the later hadronic stages has to be well understood
before any firm conclusions can be drawn. 

At SPS and RHIC/LHC energies the total hadron densities implied by the 
hadro-chemical analyses are not small, and one cannot
expect the system to decouple at this point. It has been realized, 
however, that hadronic cross sections for 
elastic scattering, especially when involving intermediate resonances 
(\eg, $\pi\pi\leftrightarrow \rho$, $\pi N \leftrightarrow \Delta$, 
$\pi K \leftrightarrow K^*$), are typically much larger than those 
for inelastic (number-changing) reactions. The associated  
timescales for thermal equilibration are thus much smaller than
for the chemical one. At SPS energies
and above, thermal equilibrium in the expanding hadron gas can be 
maintained for about 10~fm/c. This is much below typical chemical 
relaxation times and has led to the picture of a chemical freezeout 
at $(\mu_{ch},T_{ch})$ (as inferred from hadron ratios) with 
$T_{ch}\simeq$~160-180~MeV,  
followed by a thermal decoupling at significantly lower temperatures,
$T_{th}\simeq$~100-120~MeV.  
The existence of this intermediate hadronic phase is well supported 
as it importantly figures into both hadronic~\cite{Stock99} and 
electromagnetic observables~\cite{RW00}.

An immediate consequence of this picture is that all hadron species 
which are not subject to strong decays, have to be effectively conserved
in number subsequent to chemical freezeout. In statistical mechanics
language, this can be described by the build-up of (effective) 
chemical potentials, $\mu_\pi$, $\mu_K$, $\mu_\eta$, etc., in the
hadronic evolution towards thermal freezeout.  
Corresponding thermodynamic trajectories 
at SPS energies have been constructed~\cite{Bebie,SK97,HS98,PH99,RW99}.  
In this context, antibaryons play a special role. Although their 
production from AGS to RHIC energies follows chemical-freezeout 
systematics, this cannot be expected a priori, since their
annihilation cross section on baryons is large (\eg, 
$\sigma_{p\bar p\to n\pi}\simeq 50$~mb with an average $n$=5-6 at the typical 
thermal energies). 
Under SPS conditions, this issue has been resolved in 
ref.~\cite{RS01}, where it is shown that the backward reaction of 
multi-pion fusion, in connection with the rather large oversaturation 
of the pion phase space towards thermal freezeout (entailing 
$\mu_\pi^{th}\simeq 70~MeV$), is capable of approximately maintaining 
the chemical-freezeout number of antibaryons. 
On the other hand, 
with a $\bar p/p$ ratio of 5-7\%~\cite{na44,na49} and a final 
$\pi/B$ ratio (after strong decays) of around 5, the detailed treatment 
of antibaryons does not exert a large influence on the bulk 
thermodynamics of the hadronic evolution. 
The main point of this note is to show that
this is no longer true at RHIC energies. Based on the experimental
fact that also at RHIC the chemical freezeout abundances survive the
hadron gas phase, we will argue that the explicit conservation of antibaryon 
number significantly affects the composition (and possibly evolution)
of the hadronic expansion at collider energies.  In particular, 
we will construct an  explicit isentropic trajectory in the $\mu_B$-$T$ 
plane including effective number-conservation constraints.   

Before we proceed with a more quantitative assessment, let us 
further motivate our objective by the following estimate:  
RHIC data at $\sqrt{s}=130$~AGeV for central Au-Au 
collisions~\cite{phobos00,phenix01a,star01a} imply a 
midrapidity density of charged pions plus kaons of about 
$dN_{ch}/dy\simeq 600$, so that $dN_{\pi+K}/dy\simeq 900$. 
On the other hand, the total baryon number has been estimated 
to be $dN_B/dy\simeq 100$~\cite{Tser02,phenix01b}
so that $dN_{B+\bar B}/dy\simeq 165$ 
(using $\bar B/B\simeq 0.65$~\cite{star01b,phenix01b}).   
Since a (anti-) baryon carries about twice the   
entropy than that of pion, it follows that the former account for at 
least 25\% of the total entropy at midrapidity.  

Along the lines of standard hadro-chemical analyses we base the following 
calculations for the thermodynamic state variables on
an ideal hadron gas\footnote{for a discussion of the influence 
of hadronic in-medium effects, see, \eg, refs.~\cite{MFB01,Zsch01}.} 
including a set of resonance states comprised of 37 meson and 37 baryon 
species (up to masses $m_M=1.7$~GeV and $m_B=2~GeV$, 
respectively).  We will focus on conditions representative
for midrapidities in central Au+Au collisions at $\sqrt{s}=200$~AGeV
(very similar results emerge for a net-baryon free environment 
appropriate for LHC). 
For definiteness, we employ a specific entropy (\ie, entropy per
{\em net} baryon) of $S/N_B^{net}=250$ which, together with a chemical 
freezeout temperature of $T_{ch}=180$~MeV, results in a baryon chemical 
potential of $\mu_B^{ch}=24$~MeV. This agrees well  with (moderate) 
extrapolations of very similar thermal model analyses~\cite{pbm01,xu02} 
based on RHIC data at $\sqrt{s}=130$~AGeV.      
By definition, all meson-chemical potentials are zero at $T_{ch}$, 
and we also neglect (small) isospin ($\mu_I$) and strangeness ($\mu_s$)
chemical potentials. As usual, the effective pion number, which is
subject to conservation, is defined by including all strongly
decaying resonances with lifetimes shorter than the duration of
the interacting hadronic phase, \ie, 
\beq
N_\pi^{eff}= V_{FB} \sum\limits_i N_\pi^{(i)} \varrho_i(T,\mu_i) \ ,  
\label{Npi}
\eeq 
and likewise for $K$, $\eta$ and $\eta'$. In eq.~(\ref{Npi}), 
$V_{FB}$ denotes the fireball 3-volume (which is related to the 
centrality of the collision but does not affect thermodynamic 
quantities) and $\varrho_i$ the number
density of hadron $i$. The effective pion-number $N_\pi^{(i)}$ 
of a given resonance is determined by the number of pions
in its decay modes including the branching ratios, 
\eg, $N_\pi^{(\rho)}=2$, $N_\pi^{(\Delta)}=N_\pi^{(K^*)}=1$, 
$N_\pi^{(N(1520))}=0.55*1+0.45*2$. The same weighting
applies to the corresponding effective chemical potential of
each resonance, \eg, 
$\mu_\rho=2\mu_\pi$, $\mu_\Delta=\mu_N+\mu_\pi$, 
$\mu_{K^*}=\mu_\pi+\mu_K$, $\mu_{N(1520)}=\mu_N+1.45\mu_\pi$,  
which means that the strong decays and formations
are in relative chemical equilibrium.   
For a few states this distinction is not very sharp, most notably
$\omega$ and $\phi$ mesons. Their free lifetimes are well above
hadronic fireball lifetimes, but in-medium effects are expected
to alter this situation~\cite{OR01,Ra01,SW01,PM01,AK02,Ra02}. 
In the former case, some of the $\omega$'s and $\phi$'s decay
after chemical freezeout without being regenerated, \ie, their number
decreases; in the latter case, their abundances stay (at least for 
a while) in chemical equilibrium which, in the case of large pion- and 
kaon-chemical potentials, could even increase their numbers.  
In our calculation we employ an intermediate scenario keeping
the individual $\omega$ and $\phi$ numbers approximately constant
(which is numerically close to the chemical equilibrium assumption).    
An important point now is the treatment of antibaryons as first 
pointed out in ref.~\cite{Ra02}. We also conserve their number by 
introducing another effective chemical potential
$\mu_{\bar B}^{eff}$; this implies, \eg, 
$\mu_{\bar N}=-\mu_N+\mu_{\bar B}^{eff}$, 
$\mu_{\bar \Delta}=-\mu_N+\mu_{\bar B}^{eff}+\mu_\pi$.  
\bfg[!htb]
\vspace{-1.2cm}
\bce
\epsfig{file=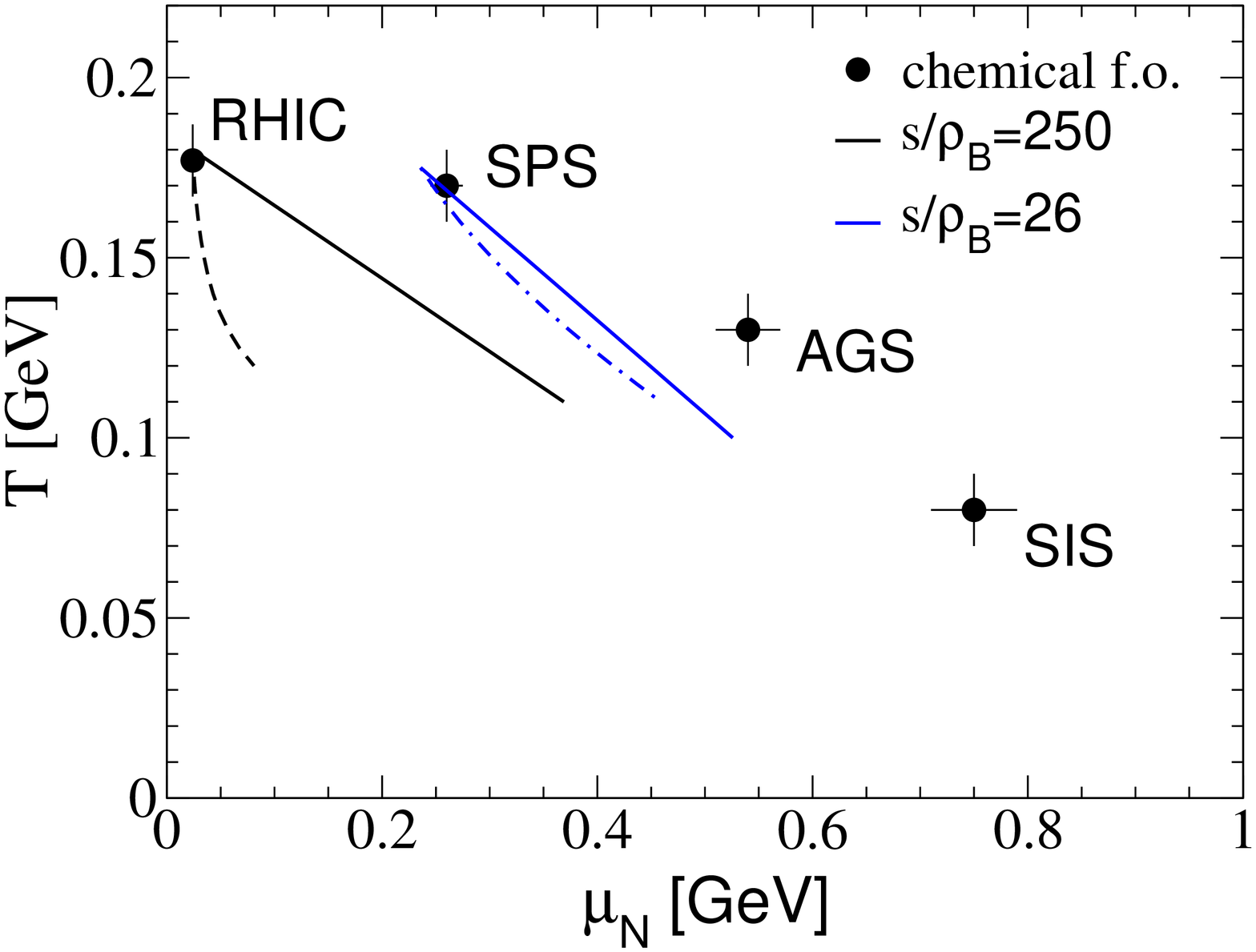,width=7cm,angle=-90}
\ece
\vspace{-0.2cm}
\caption{Isentropic thermodynamic trajectories at RHIC (left) and
SPS (right) energies, starting from the empirically determined
chemical-freezeout points~\protect\cite{pbm95,pbm96-99,CR00,pbm01}.
Full lines include antibaryon-number conservation, dashed lines do not.
(Also note that the trajectories converge towards the zero-temperature
axis at $\mu_N=m_N$.)} 
\label{fig_phasedia}
\efg
A thermodynamic trajectory in the $\mu_N$-$T$ plane is now constructed 
starting from chemical freezeout under the additional assumption 
of total entropy conservation (as applied in ideal fluid dynamics). 
Together with {\em net} baryon number conservation (which is, of course,
exact) this amounts to keeping the specific
entropy per {\em net} baryon fixed, $S/N_B^{net}= s/\varrho_B^{net}=250$
for RHIC with 
\bea
s &=& \mp  \sum\limits_i d_i \int   \frac{d^3k}{(2\pi)^3}
[\pm f(\omega_i) \ln f(\omega_i)
\label{entro}
\\
 & & \qquad \qquad + (1\mp f(\omega_i)) \ln (1\mp f(\omega_i)] \,,
%s &=& \frac{4}{3} \beta \sum\limits_i d_i \int\limits_0^\infty
%\frac{k^2dk}{2\pi^2} \ \omega_i \ f(\omega_i) \
\nonumber\\
\varrho_B^{net}&=&\sum\limits_i d_{B_i} \int\frac{d^3k}{(2\pi)^3} \ 
\left[ f^{B_i}(\mu_{B_i},T) - f^{\bar B_i}(\mu_{\bar B_i},T) \right] 
\label{rhoNnet}
\eea
denoting entropy- and net baryon number-density, respectively 
(in eq.~(\ref{entro}), upper (lower) signs refer to fermions (bosons),
and $f$ are the pertinent thermal distribution functions). 
The resulting trajectory is shown by the full line in 
Fig.~\ref{fig_phasedia}.  For comparison, we also display a trajectory 
without antibaryon-number conservation (dashed line). 
The difference is rather dramatic, indicating
a strongly increased baryon chemical potential towards the regions
of expected thermal freezeout at around $T_{th}\simeq 120$~MeV. This 
is induced by an approximately linear increase of the antinucleon
chemical potential, which is displayed in Fig.~\ref{fig_chem}  along 
with nucleon, pion, kaon, and eta chemical potentials. At chemical
freezeout the sum of baryon and antibaryon density is close to
nuclear saturation density $\rho_0=0.16$~fm$^{-3}$,   
$\varrho_{B+\bar B}\simeq 1.1\varrho_0$. On the other hand,
at $T=120$~MeV, $\varrho_{B+\bar B}\simeq 0.3 \varrho_0$ 
for the solid-line trajectory, but 
$\varrho_{B+\bar B}\simeq 0.02 \varrho_0$ for the dashed-line trajectory.   

One can readily verify that the solid-line trajectory in 
Fig.~\ref{fig_phasedia} is consistent with the
measured  $\bar p/p$ ratio, which is approximately
given by $\exp[-\Delta\mu_N/T]$ with $\Delta \mu_N=\mu_N-\mu_{\bar N}$.
It is also important to note that in the $\mu_{\bar B}^{eff}\equiv 0$ 
case, the pion-chemical potentials remain close to zero. Thus,
$\bar B$ conservation at collider
energies is at the origin of large meson chemical potentials in the
late hadronic stages\footnote{This is opposite to SPS conditions, 
where, following the arguments of ref.~\cite{RS01}, pion-oversaturation
is responsible for regenerating antibaryons.}.
The reason can be traced back to the substantial amount of entropy 
stored in  $B\bar B$ excitations. At a given temperature, the 
smaller available entropy per
pion at fixed pion number can only be realized by a finite $\mu_\pi$,
as also noted in ref.~\cite{SH92}. To illustrate this fact, we 
display in Fig.~\ref{fig_snpi} the entropy per pion in a 
$\pi$-$\rho$ gas. One finds
that for $T=120$~MeV, $S/N_\pi$ is reduced by about 20\%
when raising $\mu_\pi$ from zero to its actual value of $\sim$80~MeV
corresponding to  the solid-line RHIC-trajectory of Fig.~\ref{fig_chem}.
This is in line with the rough estimate of the entropy carried
by $B\bar B$ pairs at RHIC above.  
\bfg[!htb]
\vspace{-1.2cm}
\bce
\epsfig{file=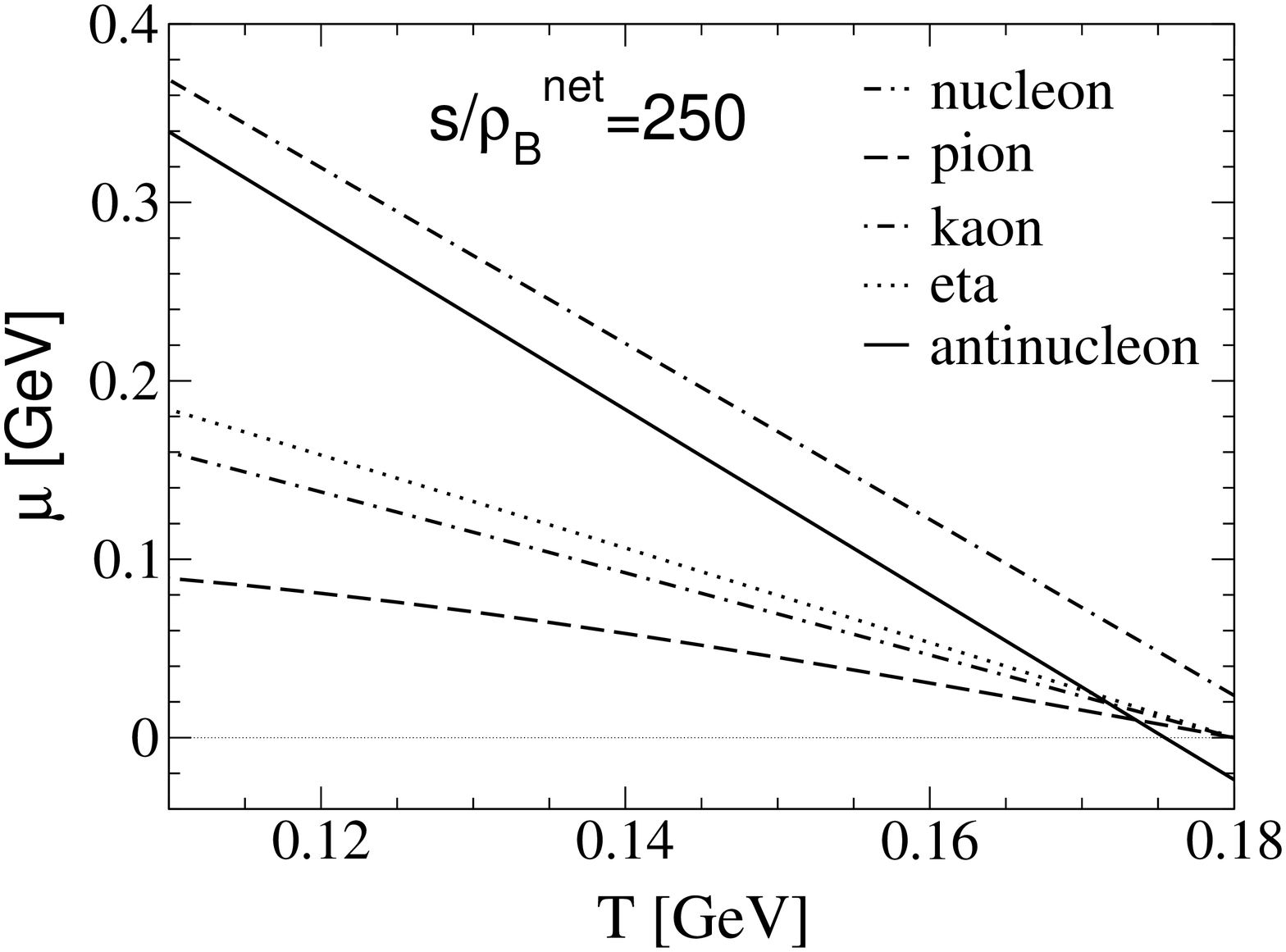,width=7cm,angle=-90}
\ece
\vspace{-0.2cm}
\caption{Temperature dependence of nucleon-,
pion-, kaon-, eta- and antinucleon-chemical potentials
for an isentropic hadronic fireball expansion at RHIC energy (200~AGeV).}
\label{fig_chem}
\efg
\bfg[!htb]
\vspace{-1.4cm}
\bce
\epsfig{file=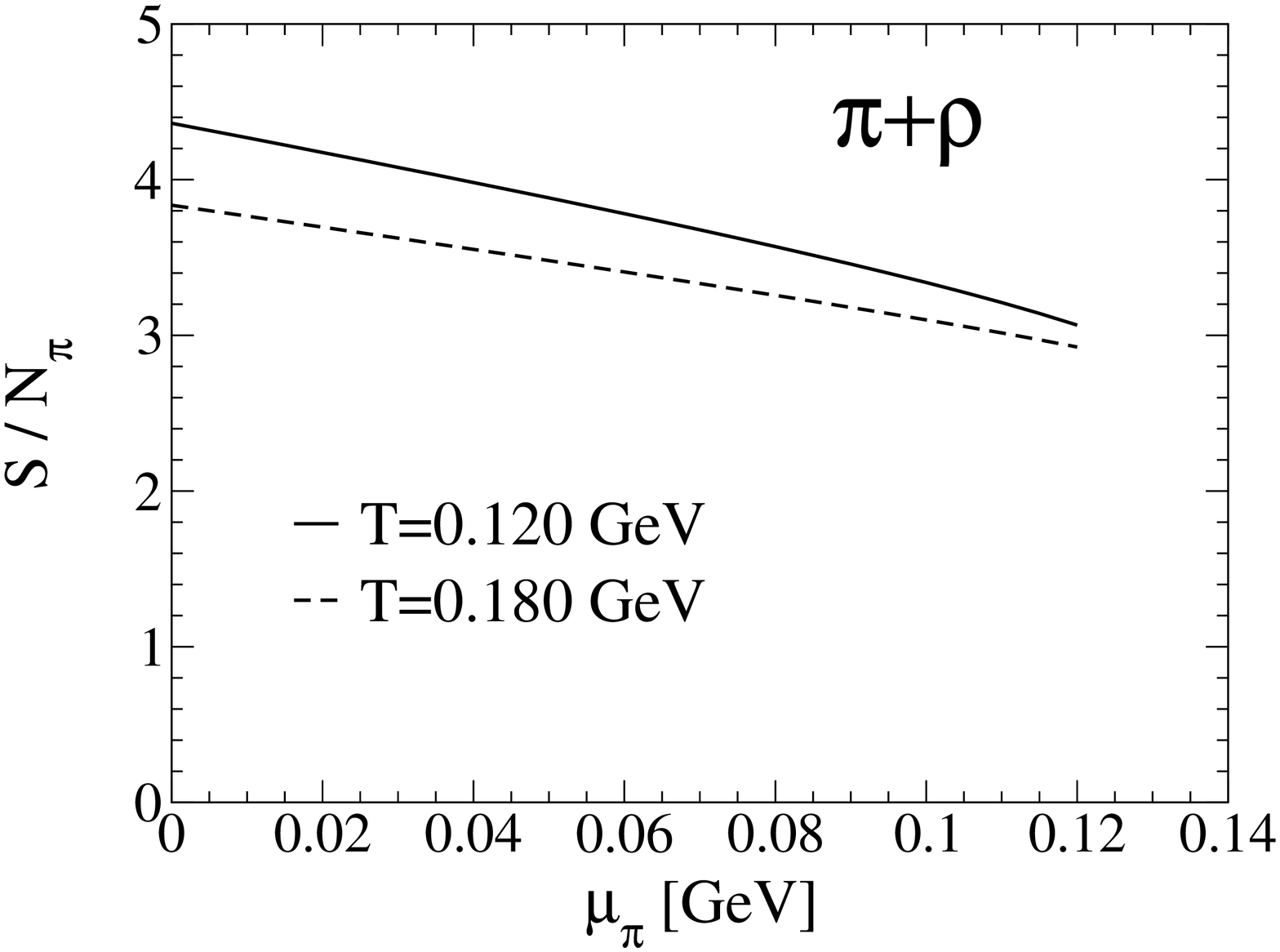,width=7cm,angle=-90}
\ece
\vspace{-0.2cm}
\caption{Entropy per pion in a $\pi$-$\rho$ gas at fixed temperature
as a function of pion-chemical potential (using $\mu_\rho=2\mu_\pi$
and counting $\rho$'s as two pions).}
\label{fig_snpi}
\efg

It furthermore follows that at given temperature the
fireball volume $V_{FB}$ of the system with finite $\mu_{\bar B}^{eff}$
is substantially reduced, \ie, the expanding hadron gas cools faster.
{\it E.g.}, at $T=120$~MeV the ratio of the 3-D fireball volumes
for the $\mu_{\bar B}^{eff}\equiv 0$ and $\mu_{\bar B}^{eff} >  0$ solution
for the trajectories amounts to a factor of $\sim$3.
Whether this will translate into smaller source size radii as
extracted via HBT correlation measurement (and thus contribute to 
an explanation of the current 'HBT-puzzle'~\cite{HK02}) remains to be
investigated within a (hydro-) dynamical simulation.
Other observable consequences of the increased densities in the
hadronic phases at RHIC concern low-mass dilepton emission, where
in-medium effects are rather sensitive to (the sum of) anti-/baryon
densities~\cite{RW00,Ra01}. Accordingly improved estimates
have been presented in ref.~\cite{Ra02}.

In our analysis we did not allude to underlying mechanisms that could 
provide a constant antibaryon abundance in the hadronic phase at RHIC. 
As mentioned above, at SPS energies this issue could be resolved~\cite{RS01}
within a thermal framework by accounting for multi-pion fusion reactions 
in the presence of a large $\mu_\pi$. Although it is conceivable 
that this mechanism is also operative at RHIC, the situation is more 
involved due to the feedback of the $\bar B$ abundance on $\mu_\pi$. 
This problem is not easily addressed within transport or cascade 
simulations either due to  
the difficulties of treating the backward reaction of multi-pion
fusion (see ref.~\cite{Cass02} for recent progress).
%(one might rather switch off $B\bar B \leftrightarrow n\pi$
%annihilation altogether).
Corresponding trajectories that have been extracted from,
\eg, UrQMD transport calculations~\cite{Brav01} are therefore
resembling the dashed lines in Fig.~\ref{fig_phasedia}.

In summary, based on the notion of hadro-chemical freezeout,
we have constructed (isentropic) thermodynamic trajectories for the 
hadronic phase in ultra-relativistic heavy-ion collisions. 
Consistency with the observed particle ratios has been enforced
by effective conservation laws, implemented via finite 
chemical potentials for stable hadrons ({\it w.r.t.}~strong 
interactions). The main emphasis was on consequences of conserved 
antibaryon-number.  At SPS energies and below, these are small, but  
at collider energies (RHIC and LHC) we have found appreciable 
modifications in the composition of the expanding hadronic system. 
In particular, the entropy 
carried by $B\bar B$ pairs is not small and has been identified as 
the origin of large meson-chemical potentials, which in turn  
leads to smaller and thus denser systems at temperatures
prior to thermal freezeout. This could have important consequences
for hadronic and electromagnetic observables, such as source radii
and medium effects in low-mass dilepton production. 
The former are rather sensitive to the detailed expansion dynamics
of a heavy-ion reaction, which we did not address here. For the
same reason, the question {\em where} on the given trajectories thermal 
freezeout in the hadronic phase occurs has not been answered, but
needs to be investigated in hydrodynamical and/or microscopic transport 
simulations.

\vskip0.2cm
 
%\centerline {\bf ACKNOWLEDGMENTS}
This work was supported by the U.S. Department of Energy 
under Grant No. DE-FG02-88ER40388.

\end{narrowtext}
\end{document}